%% file: basf2.tex
\documentclass[twocolumn]{svjour3}
\usepackage[utf8]{inputenc}
\usepackage{graphicx}
\usepackage{amsmath}
\usepackage{listings}
\usepackage{xcolor}

\usepackage{subcaption}
\usepackage{hyperref} 

\hypersetup
{
pdftitle = {basf2},
pdfkeywords = {Belle II, basf2, },
pdfproducer = {pdflatex},
}

\usepackage{microtype}

\title{The Belle II Core Software}

\author{T. Kuhr \and C. Pulvermacher \and M. Ritter \and T. Hauth \and N. Braun \\
\\
\textit{Belle II Framework Software Group}}

\institute{T. Kuhr \and M. Ritter \at
              Ludwig-Maximilians-Universit\"at, Munich, Germany
           \and
           C. Pulvermacher \at
              High Energy Accelerator Research Organization (KEK), Tsukuba, Japan
           \and
           N. Braun \and T. Hauth \at
              Karlsruhe Institute of Technology, Karlsruhe, Germany
}

\date{Received: date / Accepted: date}

\begin{document}

\maketitle
\begin{abstract}
Modern high-energy physics (HEP) enterprises, such as the Belle II experiment~\cite{b2tdr,b2tip} at the KEK laboratory in Japan, create huge amounts of data.
Sophisticated algorithms for simulation, reconstruction, visualization, and analysis are required to fully exploit the potential of these data.

We describe the core components of the Belle II software that provide the foundation for the development of complex algorithms and their efficient application on large data sets.

\end{abstract}

\section{Belle II Analysis Software Framework}
\label{sec:basf2}
\subsection{Code Structure}
\input{code}

\subsection{Basf2 Development Infrastructure and Procedures}
\input{development}

\subsection{Modules, Parameters, and Paths}
\input{modules}

\subsection{Data Store and I/O}
\input{datastore}

\subsection{Event Data Model}
\input{event_data_model}

\section{Central Services}
\subsection{Python Interface and Jupyter Notebooks}
\input{python_interface}

\subsection{Parallel Processing}
\input{parallel_processing}

\subsection{Random Numbers}
\input{random}

\subsection{Conditions Data}
\input{conditions}

\subsection{Geometry and Magnetic Field}
\input{geometry}

\section{Conclusions}
Ten years of development work with emphasis on software quality have culminated in a reliable software framework for the Belle II collaboration that is easy to use and extend with new or improved algorithms.
It fulfills the requirements for data taking, simulation, reconstruction, and analysis.
The success is illustrated by the fact that first physics results were presented to the public two weeks after collision data taking had started in Spring 2018.

While the core Belle II software is mature and robust, it must continue to accommodate the evolution of technology and requirements.
Therefore, it is crucial that expertise is preserved and carried forward to new developers, as for all other components of Belle II.

\section*{Acknowledgements}
We thank Leo Piilonen, Jan Strube, Hadrien Grasland, and Stefano Spataro for helpful comments and discussion.
We thank the KEK and DESY computing groups for valuable support.
We rely on many open-source software packages and thank the communities providing them.
We acknowledge support from BMBF and EXC153 (Germany) and MEXT (Japan).

\bibliographystyle{spphys}
\bibliography{basf2}

\end{document}

%% file: code.tex
\label{sec:code}
The core software is organized in three parts: the Belle II Analysis Software Framework \textit{basf2} containing the Belle II-specific code, the \textit{externals} containing third-party code on which basf2 depends, and the \textit{tools} containing scripts for the software installation and configuration.

\subsubsection{Basf2}
\label{sec:basf2code}
The Belle II-specific code is partitioned into about 40 packages, such as the base-level framework, one package for each detector component, the track reconstruction code, and the post-reconstruction analysis tools.
Each package is managed by one or two librarians.

The code is written in C++, with the header and source files residing in \texttt{include} and \texttt{src} subdirectories, respectively.
By default, one shared library is created per package and is installed in a top-level \texttt{lib} directory that is included in the user's library path.
The build system treats the package's contents in pre-defined subdirectories as follows: 
\begin{itemize}
\item \texttt{modules}: The code is compiled into a shared library and installed in a top-level \texttt{module} directory so that it can be dynamically loaded by basf2.
\item \texttt{tools}: C++ code is compiled into an executable and installed in a top-level \texttt{bin} directory that is included in the user's path. Executable scripts, usually written in Python, are symlinked to this directory.
\item \texttt{dataobjects}: These classes define the organization of the data that can be stored in output files. The code is linked in a shared library with the \texttt{\_data\-objects} suffix.
\item \texttt{scripts}: Python scripts are installed in a directory that is included in the Python path.
\item \texttt{data}: All files are symlinked to a top-level \texttt{data} folder.
\item \texttt{tests}: Unit and script tests (see Section~\ref{sec:development}).
\item \texttt{validation}: Scripts and reference histograms for validation plots (see Section~\ref{sec:development}).
\item \texttt{examples}: Example scripts that illustrate features of the package.
\end{itemize}

Users of basf2 usually work with centrally installed versions of basf2.
At many sites, they are provided on CVMFS~\cite{cvmfs}.
Users may also install pre-compiled binaries at a central location on their local systems with the \texttt{b2install-release} tool.
If no pre-compiled version is available for their operating system, the tool compiles the requested version from source.

\subsubsection{Externals}
\label{sec:externals}
We require a few basic packages to be installed on a system, like a compiler, make, wget, tar, and git.
The tool \texttt{b2install-prepare} checks whether these prerequisites are fulfilled and installs, if desired, the missing packages.
All other third-party code on which we rely is bundled in the externals installation.
It includes basic tools like GCC, Python 3, and bzip2 to avoid requiring a system-wide installation of specific versions at all sites, as well as HEP specific software like ROOT~\cite{root}, Geant4~\cite{geant4}, and EvtGen~\cite{evtgen}.
Some packages, like LLVM or Valgrind, are optional and not included in the compilation of the externals by default.
The number of external products has grown over time to about 60, supplemented with 90 Python packages.

The instructions and scripts to build the externals are stored in a git repository.
We use a makefile with specific commands for the download, compilation, and installation of each of the external packages.
Copies of the upstream source packages are kept on a Belle II web server to have them available, if the original source disappears.
The copies provide redundancy for the download if the original source is temporarily unavailable.
The integrity of the downloaded files is checked using their SHA 256 digests.

The libraries, executables, and include files of all external packages are collected in the common directories \texttt{lib}, \texttt{bin}, and \texttt{include}, respectively, so that each of them can be referenced with a single path.
For the external software that we might want to include in debugging efforts, such as ROOT or Geant4, we build a version with debug information to supplement the optimized version.

The compilation of the externals takes multiple hours and is not very convenient for users.
Moreover, some users experience problems because of specific configurations of their systems.
These problems and the related support effort are avoided by providing pre-compiled binary versions.
We use docker to compile the externals on several supported systems: Scientific Linux 6, Red Hat Enterprise Linux 7, Ubuntu 14.04, and the Ubuntu versions from 16.04 to 18.04.
The \texttt{b2install-externals} tool conveniently downloads and unpacks the selected version of the pre-built externals.

Because the absolute path of an externals installation is arbitrary, we have invested significant effort to make the externals location-independent.
First studies to move from the custom Makefile to Spack~\cite{spack} have been done with the aim of profiting from community solutions for the installation of typical HEP software stacks, but relocateability of the build products remains an issue.

\subsubsection{Tools}
\label{sec:tools}
The tools are a collection of shell and Python scripts for the installation and setup of the externals and basf2.
The tools themselves are set up by sourcing the script \texttt{b2setup}.
This script identifies the type of shell and then sources the corresponding sh- or csh-type setup shell script.
This script, in turn, adds the tools directory to the \texttt{PATH} and \texttt{PYTHONPATH} environment variables, sets Belle II specific environment variables, defines functions for the setup or configuration of further software components, and checks whether a newer version of the tools is available.
A pre-defined set of directories is searched for files containing site-specific configurations.
The Belle II-specific environment variables have the prefix \texttt{BELLE2} and contain information like repository locations and access methods, software installation paths, and software configuration options.

Installation of externals and basf2 releases is handled by the shell scripts \texttt{b2install-externals} and \texttt{b2install-release}, respectively.
Usually, they download and unpack the version-specific tarball of pre-compiled binaries for the given operating system.
If no binary is available, the source code is checked out and compiled.
Each version of the externals and basf2 releases is installed in a separate directory named after the version.
For the compilation of the externals, we rely on the presence of a few basic tools, like \texttt{make} or \texttt{tar}, and development libraries with header files.
Our tools contain a script that checks that these dependencies are fulfilled and, if necessary, installs the missing ones.

The command \texttt{b2setup} sets up the environment for a version-specified basf2 release.
It automatically sets up the externals version that is tied to this release, identified by the content of the \texttt{.externals} file in the release directory.
An externals version can be set up independently of a basf2 release with the \texttt{b2setup-externals} command.
The version-dependent setup of the externals is managed by the script \texttt{externals.py} in the externals directory.
Externals and basf2 releases can be compiled in optimized or debug mode using GCC.
In addition, basf2 supports the compilation with the Clang or Intel compilers.
These options can be selected with the \texttt{b2code-option} and \texttt{b2code-option-externals} commands.
A distinct subdirectory is used for the option's libraries and executables.
The commands that change the environment of the current shell are implemented as functions for sh-type shells and as aliases for csh-type shells.

The tools also support the setup of an environment for the development of basf2 code.
The \texttt{b2code-create} command clones the basf2 git repository and checks out the master branch.
The environment is set up by executing the \texttt{b2setup} command without arguments in the working directory.
If a developer wants to modify one package and take the rest from a centrally installed release, the \texttt{b2code-create} command can be used with the version of the selected release as an additional argument that is stored in the file \texttt{.release}.
The sparse checkout feature of git is used to get a working directory without checked-out code.
Packages can then be checked out individually with the \texttt{b2code-package-add} command.
The \texttt{b2setup} command sets up the environment for the local working directory and the centrally installed release.
Further tools for the support of the development work are described in Section~\ref{sec:development}.

To make it easier for users to set up an environment for the development of post-reconstruction analysis code and to encourage them to store it in a git repository, the tools provide the \texttt{b2analysis-create} command.
This requires a basf2 release version as one of the arguments and creates a working directory attached to a git repository on a central Belle II server.
The basf2 release version is stored in a \texttt{.analysis} file and used by the \texttt{b2setup} command for the setup of the environment.
The \texttt{b2analysis-get} command provides a convenient way to get a clone of an existing analysis repository and set up the build system.

The tools are designed to be able to set up different versions of basf2 and externals and thus must be independent of them.
For this reason, all binary code is placed in the externals.
When GCC and Python were embedded in the tools originally to avoid duplication in multiple externals versions, this proved difficult to manage during updates.
One of the prime challenges that we overcame in the development of the tools was to cope with the different shell types and various user environment settings.

%% file: development.tex
\label{sec:development}
The basf2 code is maintained in a git repository.
We use Bitbucket Server~\cite{bitbucket} to manage pull requests.
This provides us with the ability to review and discuss code changes in pull requests before they are merged to the main development branch in the git repository.
Compared to the previous workflow based on subversion, it helps the authors to improve the quality of their code and allows the reviewers to get a broader view of the software.
We exploit the integration with the Jira~\cite{jira} ticketing system for tracking and planning the development work.

Developers obtain a local copy of the code with the \texttt{b2code-create} tool (see Section~\ref{sec:tools}).
The build system is based on SCons~\cite{scons} because, compared to the HEP standard CMake, the build process is a one-step procedure and the build configuration is written in Python, a language adopted already for the basf2 configuration steering files (see Section~\ref{sec:pythoninterface}).
The time SCons needs to determine the dependencies before starting the build is reduced by optimizations, such as bypassing the check for changes of the externals.
Developers and users usually do not have to provide explicit guidance to the build system; they only have to place their code in the proper subdirectories.
However, if the code references a set of linked libraries, the developer indicates this in the associated, typically three-line, \texttt{SConscript} file.

We implement an access control for git commits to the master branch using a hook script on the Bitbucket server.
Librarians, identified by their user names in a \texttt{.librarians} file in the package directory, can directly commit code in their package.
They can grant this permission to others by adding them to a \texttt{.authors} file.
All Belle II members are permitted to commit code to any package in \texttt{feature} or \texttt{bugfix} branches.
The merging of these branches to the master via pull requests must be approved by the librarians of the affected packages.
Initially, when subversion was used for version control, direct commits to the master were the only supported workflow, but after the migration to git pull requests are the recommended and more common way of contributing.

We have established coding conventions to achieve some conformity of the code.
Because most of them cannot be enforced technically, we rely on developers and reviewers to follow them.
We do enforce a certain style to emphasize that the code belongs to the collaboration and not to the individual developer.
The AStyle tool~\cite{astyle} is used for C++ code and pep8~\cite{pep8} and autopep8~\cite{autopep8} for Python code.
Some developers feel strongly about the code formatting, and so, we make it easy to follow the rules and reduce their frustration by providing the \texttt{b2code-style-check} tool to print style violations and the \texttt{b2code-style-fix} tool to automatically fix them.
The style conformity is checked by the Bitbucket server hook upon push to the central repository.
It also rejects files larger than 1\,MB to prevent an uncontrolled growth of the repository size.
To provide feedback to developers as early as possible and to avoid annoying rejections when commits are pushed to the central repository, we implement the checks of access rights, style, and file size in a hook for commits to the local git repository.

To facilitate test-driven development, unit tests can be implemented in each package using Google Test~\cite{googletest}.
These are executed with the \texttt{b2test-units} command.
Test steering files in all packages can be run with the \texttt{b2test-scripts} command.
It compares the output to a reference file and complains if they differ or if the execution fails.
The unit and steering file tests are executed by the Bamboo~\cite{bamboo} build service, whenever changes are pushed to the central repository.
Branches can only be merged to the master if all tests succeed.

The tests are also executed by a Buildbot~\cite{buildbot} continuous integration system that compiles the code with the GCC, Clang, and Intel compilers and informs the authors of commits about new errors or warnings.
Once a day, the Buildbot runs Cppcheck, a geometry overlap check, Doxygen and Sphinx~\cite{sphinx} documentation generation, and a Valgrind memory check.
The results are displayed on a web page, and the librarians are informed by email about issues in their packages.
A detailed history of issues is stored in a MySQL database with a web interface that also shows the evolution of the execution time, output size, and memory usage of a typical job.

Higher-level quality control is provided by the validation framework.
It executes scripts in a package's \texttt{validation} subdirectory to generate simulated data files and produce plots from them.
The validation framework then spawns a web server to display the plots in comparison with a reference as well as results from previous validation runs.
A software quality shifter checks the validation plots produced each night for regressions and informs the relevant individual(s) if necessary.

As a regular motivation for the libarians to review the changes in their package, we generate monthly builds.
For a monthly build, we require all librarians to agree on a common commit on the master branch.
They signal their agreement using the \texttt{b2code-package-tag} command to create a git tag of the agreed-upon common commit with a tag name composed of the package name and a version number.
The command asks for a summary of changes that is then used as tag message and included in the announcement of the monthly build.
The procedure of checking the agreement, building the code, and sending the announcement is fully automated with the Buildbot.

An extensive manual validation, including the production of much larger data samples, is done before releasing a major official version of basf2.
Based on these major versions, minor or patch releases that require less or no validation effort are made.
In addition, light basf2 releases containing only the packages required to analyze mini DST (mDST, see Section~\ref{sec:edm}) data can be made by the analysis tools group convener.
This allows for a faster release cycle of analysis tools.
Each release is triggered by pushing a tag to the central repository.
The build process on multiple systems and the installation on CVMFS is then automated.

In maintaining or modifying the development infrastructure and procedures, we aim to keep the thresholds to use and contribute to the software as low as possible and, at the same time, strengthen the mindset of a common collaborative project and raise awareness of code quality issues.
This includes principles like early feedback and not bothering developers with tasks that can be done by a computer.
For example, the tools complain about style-rule violations already on commits to the local git repository and offer programmed corrections.
In this way, users and developers can focus on the development of their code and use their time more efficiently.

%% file: modules.tex
\label{sec:modules}
The data from the Belle II detector, or simulations thereof, are organized into a set of variable-duration runs, each containing a sequence of independent events.
An event records the measurements of the by-products of an electron-positron collision or a cosmic ray passage.
A set of runs with similar hardware state and operational characteristics is classified as an experiment.
Belle II uses unsigned integers to identify each experiment, run, and event.

The basf2 framework executes a series of dynamically loaded modules to process a collection of events.
The selection of modules, their configuration, and their order of execution are defined via a Python interface (see Section~\ref{sec:pythoninterface}).

A module is written in C++ or Python and derived from a \texttt{Module} base class that defines the following interface methods:
\begin{itemize}
\item \texttt{initialize()}: called before the processing of events to initialize the module.
\item \texttt{beginRun()}: called each time before a sequence of events of a new run is processed, e.g., to initialize run-dependent data structures like monitoring histograms.
\item \texttt{event()}: called for each processed event.
\item \texttt{endRun()}: called each time after a sequence of events of the same run is processed, e.g., to collect run-summary information.
\item \texttt{terminate()}: called after the processing of all events.
\end{itemize}
Flags can be set in the constructor of a module to indicate, for example, that it is capable of running in parallel processing mode (see Section~\ref{sec:parallel}).
The constructor sets a module description and defines module parameters that can be displayed on the terminal with the command \texttt{basf2 -m}.

A module parameter is a property, whose value (or list of values) can be set by the user at runtime via the Python interface to tailor the module's execution.
Each parameter has a name, a description, and an optional default value.

The sequence in which the modules are executed is stored in an instance of the \texttt{Path} class.
An integer result value that is set in a module's \texttt{event()} method can be used for a conditional branching to another path.
The processing of events is initiated by calling the \texttt{process()} method with one path as argument.
The framework checks that there is exactly one module that sets the event numbers.
It also collects information about the number of module calls and their execution time.
This information can be printed after the event processing or saved in a ROOT file.

Log messages are managed by the framework and can be passed to different destinations, like the terminal or a text file, via connector classes.
Methods for five levels of log messages are provided:
\begin{itemize}
\item \texttt{FATAL}: for situations, where the program execution cannot be continued.
\item \texttt{ERROR}: for things that went wrong and must be fixed. If an error happens during initialization, event processing is not started.
\item \texttt{WARNING}: for potential problems that should not be ignored and only accepted if understood.
\item \texttt{INFO}: for informational messages that are relevant to the user.
\item \texttt{DEBUG}: for everything else, intended solely to provide useful detailed information for developers. An integer debug level is used to control the verbosity.
\end{itemize}
The log and debug levels can be set globally, per package, or per module.

%% file: datastore.tex
\subsubsection{Data Store}
\label{sec:datastore}

Modules exchange data via the \textit{Data Store} that provides a globally accessible interface to mutable objects or arrays of objects.
Objects (or arrays of objects) are identified by name that, by default, corresponds to the class name.
By convention, arrays are named by appending an ``s'' to the class name.
Users may choose a different name to allow different objects of the same type simultaneously.
Objects in the Data Store can have either permanent or event-level durability.
In the latter case, the framework clears them before the next data event is processed.
Client code can add objects to the Data Store, but not remove them.

Within one event, two distinct arrays of objects in the Data Store can have weighted many-to-many relations between their elements.
For example, a higher-level object might have relations to all lower-level objects that were used to create it.
Each relation carries a real-valued weight that can be used to attach quantitative information such as the fraction a lower-level object contributed to the higher-level one.
The relationship information is stored in a separate object; no direct pointers appear in the related objects.
This allows us to strip parts of the event data, without affecting data integrity: if one side of a relationship is removed, the whole relation is dropped.
The relations are implemented by placing a \texttt{RelationArray} in the Data Store that records the names of the arrays it relates, as well as the indices and weights of the related entries.
As the Data Store permits only appending entries to an array, the indices are preserved.
The name of the relations object is formed by placing ``To'' between the names of the related arrays.

The interface to objects in the Data Store is implemented in the templated classes \texttt{StoreObjPtr} for single objects and \texttt{StoreArray} for arrays of objects, both derived from the common \texttt{StoreAccessorBase} class.
They are constructed with the name identifying the objects; without any argument, the default name is used.
Access to the objects is type-safe and transparent to the event-by-event changes of the Data Store content.
To make the access efficient, the \texttt{StoreAccessorBase} translates the name on first access to a pointer to a \texttt{DataStoreEntry} object in the Data Store.
The \texttt{DataStoreEntry} object is valid for the lifetime of the job and contains a pointer to the currently valid object, which is automatically updated by the Data Store.
Access to an object in the Data Store thus requires an expensive string search only on the first access, and then a quick double dereferencing of a pointer on subsequent accesses.

The usage of relations is simplified by deriving the objects in a Data Store array from \texttt{RelationsObject}.
It provides methods to directly ask an object for its relations to, from, or with (ignoring the direction) other objects.
Non-persistent data members of \texttt{RelationsObject} and helper classes are used to make the relations lookup fast by avoiding regeneration of information that was obtained earlier.

We provide an interface to filter, update or rebuild relations when some elements are removed from the Data Store.
It is possible to copy whole or partial arrays in the Data Store, where new relations between the original and copied arrays are created, and, optionally, the existing relations of the original array are copied.

\subsubsection{I/O}
We use ROOT for persistency.
This implies that all objects in the Data Store must have a valid ROOT dictionary.
The \texttt{RootOutputModule} writes the content of the Data Store with permanent and event durability to a file with two separate \texttt{TTree}s, with a branch for each Data Store entry.
The selection of branches, the file name, and some tree configurations can be specified using module parameters.
The corresponding module for reading ROOT files is the \texttt{RootInputModule}.

The \texttt{RootOutputModule} writes an additional object named \texttt{FileMetaData} to the permanent-durability tree of each output file.
It contains a logical file name, the number of events, information about the covered experiment/run/event range, the steering file content, and information about the file creation.
The file metadata also contains a list of the logical file names of the input files, called parents, if any.

This information is used for the index file feature.
A \texttt{RootInputModule} can be asked to load, in addition to the input file, its ancestors up to a generational level given as a parameter.
A file catalog in XML format, created by the \texttt{RootOutputModule}, is consulted to translate logical to physical file names for the ancestor files.
The unique event identifier is then used to locate and load the desired event.
With the index file feature, one can produce a file containing only \texttt{EventMetaData} objects (see next section) of selected events, and then use this as the input file in a subsequent job to access the selected events in its parents.
File-reading performance is not optimal, however, since the usual structure of \texttt{TTree}s in ROOT files is not designed for sparse event reading.
The index file feature can be used also to add objects to an existing file without copying its full content or to access lower level information of individual events for display or debug purposes.

The Belle II data-acquisition system uses a custom output format with a sequence of serialized ROOT objects to limit the loss of events in case of malfunctions.
The files in this format are transient; they are converted to standard ROOT files for permanent storage.

%% file: event_data_model.tex
\label{sec:edm}
The Data Store implementation makes no assumption about the event data model.
It can be chosen flexibly to match specific requirements.
In basf2, the full event data model is defined dynamically by the creation of objects in the Data Store by the executed modules.
The only mandatory component is the \texttt{EventMetaData} object.
It uniquely identifies an event by its event, run, and experiment numbers and a production identifier to distinguish simulated events with the same event, run, and experiment numbers.
The other data members store the time when the event was recorded or created, an error flag indicating problems in data taking, an optional weight for simulated events, and the logical file name of the parent file for the index file feature.

The format of the raw data is defined by the detector readout.
Unpacker modules for each detector component convert the raw data to digit objects.
In case of simulation, the digit objects are created by digitizer modules from energy depositions that are generated by Geant4 and stored as detector-specific \texttt{SimHits}.
The use of a common base class for \texttt{SimHits} allows for a common framework to add energy depositions from simulated machine-induced background to that of simulated physics signal processes.
This is called background mixing.

The output of the reconstruction consists mainly of detector-specific objects.
In contrast, the \texttt{RecoTrack} class is used to manage the pattern recognition and track fitting across multiple detectors.
It allows us to add hits to a track candidate and is interfaced to GENFIT~\cite{genfit,genfit2} for the determination of track parameters.

The subset of reconstruction \texttt{dataobjects} used in physics analyses, called mini data summary table (mDST), is explicitly defined in the steering file function \texttt{add\-\_mdst\-\_output}.
It consists of the following classes:
\begin{itemize}
\item \texttt{Track}: the object representing a reconstructed trajectory of a charged particle, containing references to track fit results for multiple mass hypotheses and a quality indicator that can be used to suppress fake tracks.
\item \texttt{TrackFitResult}: the result of a track fit for a given particle hypothesis, consisting of five helix parameters, their covariance matrix, a fit $p$-value, and the pattern of layers with hits in the vertex detector and drift chamber.
\item \texttt{V0}: candidate of a $K^0_S$ or $\Lambda$ decay or of a converted photon, with references to the pair of positively and negatively charged daughter tracks and track fit results. The vertex fit result is not stored as it can be reconstituted at analysis level.
\item \texttt{PIDLikelihood}: the object that stores, for a charged particle identified by the related track, the likelihoods for being an electron, muon, pion, kaon, proton or deuteron from each detector providing particle identification information.
\item \texttt{ECLCluster}: reconstructed cluster in the electromagnetic calorimeter, containing the energy and position measurements and their correlations, along with shower-shape variables; a relation is recorded if the cluster is matched to an extrapolated track.
\item \texttt{KLMCluster}: reconstructed cluster in the $K^0_L$ and muon (KLM) detector, providing a position measurement and momentum estimate with uncertainties; a relation is recorded if the cluster is matched to an extrapolated track.
\item \texttt{KlId}: candidate for a $K^0_L$ meson, providing particle identification information in weights of relations to KLM and/or ECL clusters.
\item \texttt{TRGSummary}: information about level 1 trigger decisions before and after prescaling, stored in bit patterns.
\item \texttt{SoftwareTriggerResult}: the decision of the high-level trigger, implemented as a map of trigger names to trigger results.
\item \texttt{MCParticle}: the information about a simulated particle (in case of simulated data), containing the momentum, production and decay vertex, relations to mother and daughter particles, and information about traversed detector components; relations are created if simulated particles are reconstructed as tracks or clusters.
\end{itemize}

The average size of an mDST event is a critical performance parameter for the storage specification and for the I/O-bound analysis turnaround time.
Therefore, the mDST content is strictly limited to information that is required by general physics analyses.
In particular, no raw data information is stored.
For detailed detector or reconstruction algorithm performance studies as well as for calibration tasks a dedicated format, called cDST for calibration data summary table, is provided.

%% file: python_interface.tex
\subsubsection{Python Interface}
\label{sec:pythoninterface}

To apply the functionality described in \autoref{sec:basf2} to a data processing task -- at the most basic level, arranging appropriate modules into a path and starting the event processing --
basf2 provides a Python interface.
Typically, users perform tasks using Python scripts (called ``steering files'' in this context), but interactive use is also supported.
Figure~\ref{fig:steering} shows a minimal example for the former, while \autoref{sec:jupyternotebooks} discusses applications for the latter.

\begin{figure}[htbp]
\begin{lstlisting}
#!/usr/bin/env python3
# -*- coding: utf-8 -*-

# Generate 100 events with event numbers 0 to 99 that contain only the event meta data.

import basf2
main = basf2.create_path()
main.add_module('EventInfoSetter', evtNumList=[100])
basf2.process(main)

\end{lstlisting}
\caption{Example of a basf2 steering file.}
\label{fig:steering}
\end{figure}

Python is a very popular language and provides an easy-to-understand syntax that new users can rather quickly deploy to use the framework efficiently.
It allows us to harness the power of a modern scripting language for which copious (third-party) packages are available.
We exploit this, for example, to build a higher-level framework for performing typical analysis tasks in a user-friendly way.
The docstring feature of Python is used to generate documentation web pages with Sphinx.

We use Boost.Python~\cite{boost.python} to expose the basf2 framework features in Python.
While steering files can be executed by passing them directly to the Python interpreter, we also provide the \texttt{basf2} executable as an alternative to add framework-specific command line arguments.
Among these are options to print versioning information, list available modules and their description, and specify input or output file names.

Besides the implementation of modules in C++, the framework allows the user to execute modules written in Python.
This makes it even easier for users to write their own module code because it can be embedded in the steering file.
It can also facilitate rapid prototyping.
Even so, the modules provided by the framework are written in C++ (with a few exceptions for tasks that are not performance critical) to profit from the advantages of compiled code.

Using PyROOT~\cite{pyroot}, Python access to the Data Store is provided by classes resembling the \texttt{StoreObjPtr} and \texttt{StoreArray} interfaces.
In an equivalent way, interface classes provide access to conditions data, such as calibration constants (see Section~\ref{sec:conditions}).

A feature that facilitates development and debugging is the possibility to interrupt the event processing and present an interactive Python prompt.
In the interactive session based on IPython~\cite{ipython}, the user can inspect or even modify the processed data.

\subsubsection{Jupyter Notebooks}
\label{sec:jupyternotebooks}
Typical HEP user-level analyses for processing large data samples are mostly based on the execution of small scripts written in Python or ROOT macros that call complex compiled algorithms in the background.
Jupyter notebooks~\cite{jupyter} allow a user to develop Python-based applications that bundle code, documentation and results (such as plots).
They provide an enriched browser-based working environment that is a front-end to an interactive Python session that might be hosted centrally on a remote high-performance computing cluster.
Jupyter notebooks include convenient features like syntax highlighting and tab-completion as well as integration with data-analysis tools like ROOT, matplotlib~\cite{matplotlib} or pandas~\cite{pandas}.

The integration of Jupyter into basf2 simplifies the process of creating and processing module paths within Jupyter notebooks and represents a natural next step beyond the integration of Python into basf2. 
The package for the interplay between Jupyter and basf2 is encapsulated into a basf2-agnostic hep-ipython-tools project~\cite{hep-ipython-tools} that can be used with the framework code of other experiments.

The processing of one or more paths is decoupled into an abstract \textit{calculation} object, which plays well with the interactivity of the notebooks, because multiple instances of this calculation can be started and monitored, while continuing the work in the notebook.
Abstracting the basf2 calculation together with additional interactive widgets and convenience functions for an easier interplay between Juypter and basf2 not only improves the user experience, but also accentuates the narrative and interactive character of the notebooks.

The decoupling of the calculations is achieved using the multiprocessing library and depends heavily on the ability to steer basf2 completely from the Python process. 
Queues and pipelines are used from within the basf2 modules to give process and runtime-dependent information back to the notebook kernel.
The interactive widgets are created using HTML and JavaScript and display information on the modules in a path, the content of the data store or the process status and statistics.

%% file: parallel_processing.tex
\label{sec:parallel}
For the past several years, the processing power of CPUs has grown by increasing the number of cores instead of the single-core performance.
To efficiently use modern CPU architectures, it is essential to be able to run applications on many cores.

The trivial approach of running multiple applications, each using one core, neglects the sharing of many other resources.
In particular, the size of and the access to the shared memory can be bottlenecks.
The amount of memory per core on typical sites used by HEP experiments has remained in the range of 2-3 GB for many years.

A more efficient shared use of memory can be achieved by multithreaded applications.
The downside is, that this imposes much higher demands and limitations on the code to make it thread safe.
While the development of thread-safe code can be assisted by libraries, it requires a non-trivial change in how code is written.
Few, if any, Belle II members have the skills to write thread-safe code.
Developing a multithreaded framework would require educating on the order of a hundred developers.
Additionally, the multiprocessing savings are sufficient for stable operation of the Belle~II software, as the memory consumption of a single event is small.

\begin{figure}[htbp]
  \centering
  \includegraphics[width=0.45\textwidth]{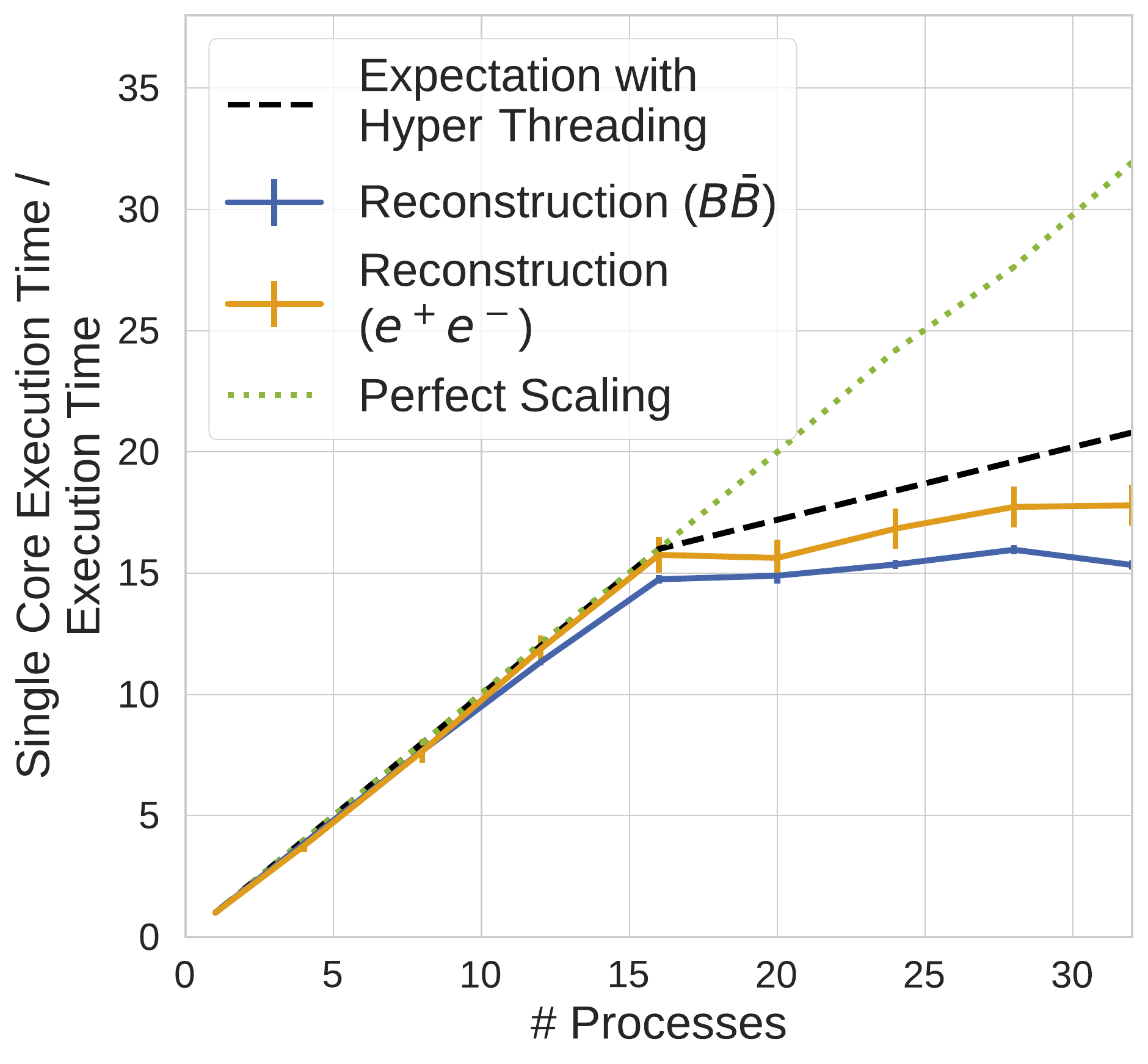}
  \caption{Scaling of parallel processing rate vs. number of parallel processes measured on a 16-core machine for smaller ($e^+e^-$) and larger ($B\bar{B}$) events. As reference, the expected 
perfect scaling is plotted as the dotted line, assuming a 20\% gain in the hyper-threading domain.
The measured speedup when using sleep instructions is plotted with a dotted green line.}
  \label{fig:parallel_processing}

  \includegraphics[width=0.45\textwidth]{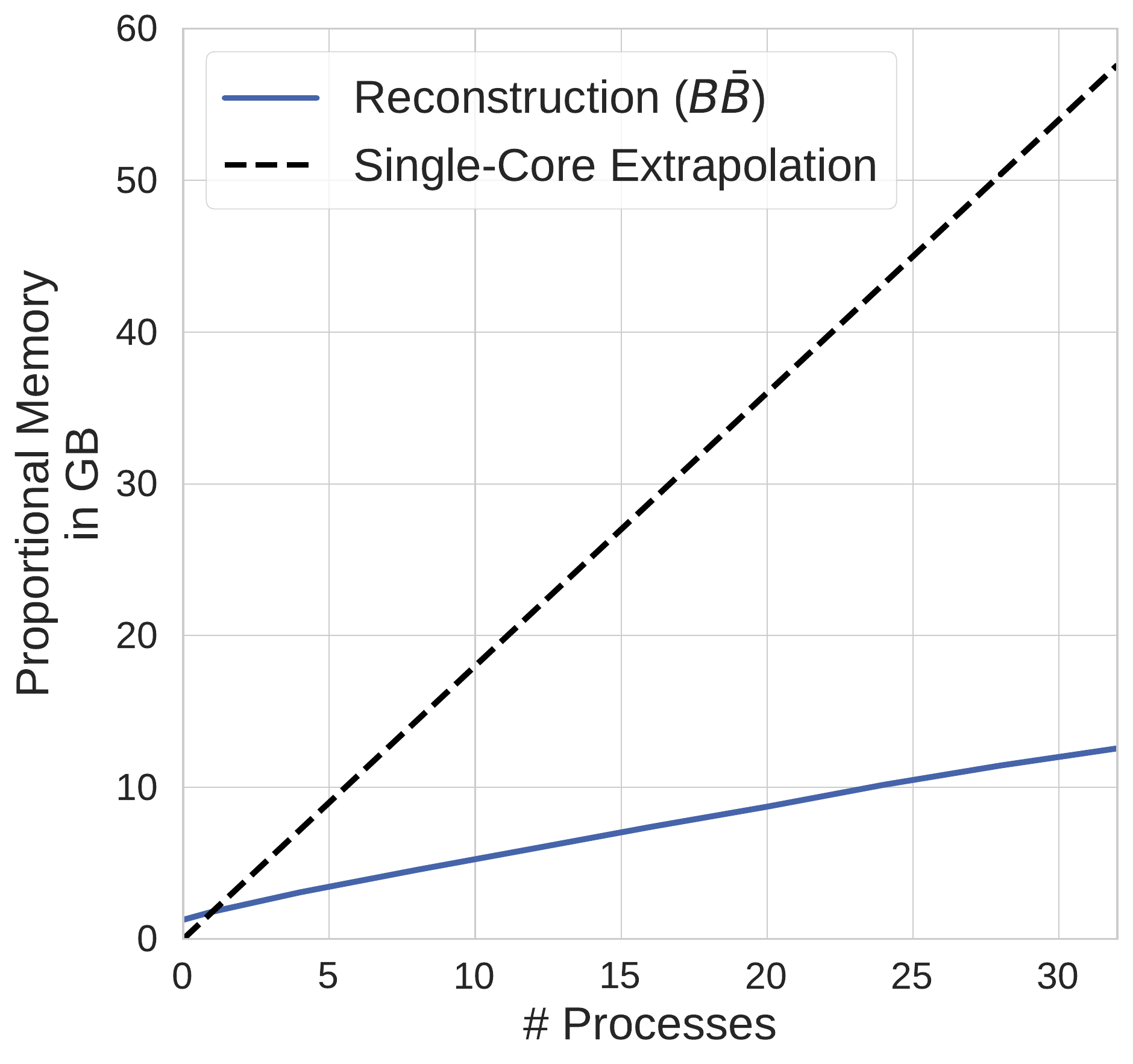}
  \caption{Proportional memory usage of parallel processing jobs for $B\bar{B}$ events. The graph for $e^+e^-$ events is very similar. For comparison, the memory usage of a single-core job times the number of processes is plotted as the dashed line.}
  \label{fig:parallel_processing_memory}
\end{figure}

In our solution, we have implemented a parallel processing feature, where processes are started by forking and each of them processes the data of a separate complete event.
As the processes have independent memory address spaces, developers do not have to care about thread-safe data access.
Still, we can significantly reduce the memory consumption of typical jobs because of the copy-on-write technology used by modern operating systems.
A large portion of the memory is used for the detector geometry.
Because it is created before the forking and does not change during the job execution, multiple processes share the same geometry representation in memory.
Figure~\ref{fig:parallel_processing} illustrates the scaling of a basf2 job's execution time with the number of parallel processes on a 16-core machine.
For both event reconstruction scenarios, one with smaller ($e^+e^-$) and the other with larger ($B\bar{B}$) event sizes, the scaling is either equal 
or very close to the theoretical linear expectation until the number of parallel processes exceeds the number of cores.
The minor loss in efficiency when the number of processes reaches the number of cores can be attributed to shared resources, like level-3 caches, used by all processing cores.
The memory saving is illustrated in Figure~\ref{fig:parallel_processing_memory}.

Each module indicates via a flag (see Section~\ref{sec:modules}) to the framework, whether it can run in parallel processing mode, or not.
Notably, the input and output modules that read or write ROOT files cannot.
As the input and output modules are usually at the beginning and end of a path, respectively, the framework analyzes the path and splits it into three sections.
The first and last section are each executed in a single process.
Only the middle section is executed in multiple processes.
The beginning of the middle section is defined by the first module that can run in parallel processing mode.
The next module that is not parallel processing capable defines the beginning of the third section.
Each event is processed by all three sections, but only by one process at any given time.
After the first section is completed the event is passed to exactly one worker process which in turn sends it to the third section.

To transfer the event data among these processes, dedicated transmitter and receiver modules are added at the end or beginning of the sections.
A transmitter module serializes the event data using the streamers generated by ROOT and writes it to a ring buffer in shared memory.
A receiver module reads the event data and deserializes it, so that it becomes available in the Data Store of the process.
In case of a run transition the input process waits until all receiving processes are finished to avoid mixing of events from different runs in the output.
The interprocess communication is based on System V shared memory.
A replacement of the custom solution by ZeroMQ~\cite{zeromq} is being evaluated.

This parallel processing scheme works well if the computational effort of the modules in the middle section dominates over the input, output, and (de)serialization load.
For high-throughput jobs with little computational demands, the serialization and deserialization impose a sizable penalty, so that the multiple cores of a CPU are not optimally exploited. 
For typical Belle II reconstruction jobs and event data sizes, we have verified with up to 20 concurrent processes, which is well within 
the envelope of parallelism we currently foresee to deploy during the online reconstruction or grid simulation and reconstruction, that the input and output processes do not become a bottleneck.

%% file: random.tex
Belle~II will generate very large samples of simulated data for a broad array of physics processes to provide signal and background expectations with a precision that is much better than available in real data.
We have to ensure that this production is not
hindered by issues with the pseudorandom number generator (PRNG). A PRNG is a
deterministic algorithm to generate numbers whose properties approximate those
of random numbers while being completely deterministic. It has an
internal state that uniquely determines both the next random number and the next
internal state. If the internal state is known at some point, all
subsequent random numbers can be reproduced.

For Belle~II, we chose xorshift1024*~\cite{DBLP:journals/corr/Vigna14}, a newer
generation PRNG based on the Xorshift algorithm proposed by
Marsaglia~\cite{Marsaglia:2003:JSSOBK:v08i14}. It generates 64-bit random numbers
with a very simple implementation, operates at high speed, and passes all
well-known statistical tests with an internal state of only 128~bytes
(1024~bits). This PRNG is used consistently throughout the framework for all
purposes: from event generation to simulation to analysis.

To ensure that events are independent, we seed the state of the
random generator at the beginning of each event using a common,
event-independent seed string together with information uniquely identifying the
event. To minimize the chance for seed collisions between different events, we
calculate a 1024~bit SHAKE256~\cite{sha3} hash from this information that we
use as the generator seed state. This also allows us to use a common seed string
of arbitrary length.

The small generator state also allows us to pass the random generator for each
event along with the event data in parallel-processing mode to achieve reproducibility independently of the
number of worker processes.

%% file: conditions.tex
\label{sec:conditions}
In addition to event data and constant values, we have a number of settings or
calibrations that can evolve over time but not on a per-event rate. These are
called ``conditions'' and their values are stored in a central Conditions Database (CDB)~\cite{conditions-db}.

Conditions are divided into payloads. Each payload is one atom of
conditions data and has one or more ``intervals of validity'' (IoV) -- the run
interval in which the payload is valid. One complete set of payloads and their
IoVs are identified by a global tag. There can be multiple global tags to provide, for
example, different calibration versions for the same run ranges. When a new global tag is
created, it is open for modifications so that assignments of IoVs to payloads can be added or removed.
Once a global tag is published, it becomes immutable.

The CDB is implemented as a representational state transfer (REST) service.
Communication is performed by standard HTTP using XML or JSON data. By design, the CDB is
agnostic to the contents of the payloads and only identifies them by
name and revision number. The integrity of all payloads is verified using a
checksum of the full content. Clients can query the CDB to obtain all payloads
valid for a given run in a given global tag.

The choice of a standardized REST API makes the client implementation
independent of the actual database implementation details and allows for a
simple and flexible implementation of clients in different programming
languages.

In addition to communication with the CDB, we have implemented a local database
backend that reads global tag information from a text file and uses the
payloads from a local folder. This allows us to use the framework without
connection to the internet, or if the CDB is unreachable or unresponsive, provided the local copies of the necessary
payloads exist. This local database is created
automatically in the working directory for all payloads that are downloaded
from the server during a basf2 job execution.

Multiple metadata and payload sources can be combined. By default, global tags
are obtained from the central server and payloads from a local database on CVMFS that is automatically updated in regular intervals.
If a payload is not found in any local folder, it is downloaded directly from the server.
If the central database is not available, the global tag is taken from the local database on CVMFS.

\subsubsection{Access of Conditions Objects}

By default, the framework assumes that payload contents are serialized
ROOT objects and manages the access to them, but direct access to payload files of any type is possible, too. User
access to conditions objects is provided by two interface classes, one for
single objects called \texttt{DBObjPtr} and one for arrays of objects
called \texttt{DBArray}. These classes reference \texttt{DBEntry} payload objects
in the \texttt{DBStore} global store. Multiple
instances of the interface class point to the same object. It is identified
by a name that is, by default, given by the class name. Access to the
conditions objects is available in C++ and in Python. The class interfaces
are designed to be as close as possible to the interface
for event-level data (see Section~\ref{sec:datastore}),
so that users can use the same concepts for both.

\begin{figure*}[htbp]
  \centering
  \includegraphics[width=\textwidth]{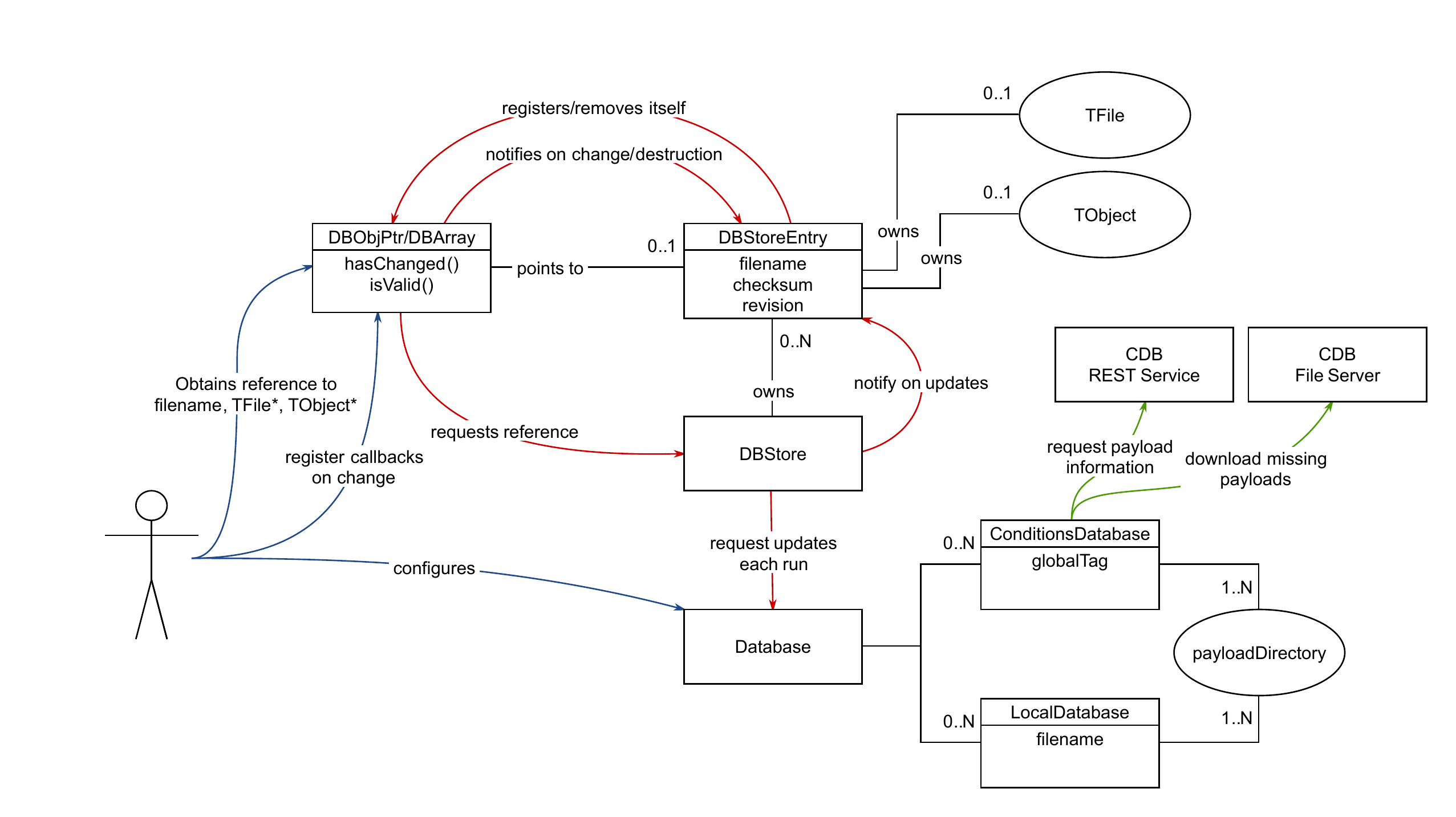}
  \caption{Relations between all entities for the Conditions Database Client.
    The user usually only interacts with the \texttt{DBObjPtr} and
    \texttt{DBArray} objects and maybe configures the database sources (shown in
    blue). Everything else is handled transparently, including the communication
    with the CDB (shown in green).}\label{fig:database-relations}
\end{figure*}

The interface classes always point to the correct payload objects for the current
run; updates are transparent to the user. If the user needs to be aware when the
object changes, they can either manually check for changes, or register a callback
function for notification. Figure~\ref{fig:database-relations}
visualizes the relations among the entities.

The CDB handles payloads at run granularity, but the framework can
transparently handle conditions that change within a run: if the payload is a
ROOT object inheriting from the base class \texttt{IntraRunDependency}, the
framework transparently checks for each event whether an update of the conditions data is required.
Different specializations of \texttt{IntraRunDependency} can be implemented: for
example, changing the conditions depending on event number or time stamp.

\subsubsection{Creation of Conditions Data}

To facilitate easy creation of new conditions data -- for example, during
calibration -- we provide two payload creation classes, \texttt{DBImportObj} and
\texttt{DBImportArray}. They have an interface very similar to \texttt{DBObjPtr}
and \texttt{DBArray}. Users
instantiate one of the creation classes, add objects to them and commit them to
the configured database with a user-supplied IoV. This includes support for
intra-run dependency. The capability to use a local file-based database allows for
easy preparation and validation of new payloads before they are uploaded to the CDB.

\subsubsection{Management of CDB Content}

To simplify the inspection and management of the CDB contents, we provide the
\texttt{b2conditionsdb} tool that uses the
requests package~\cite{python-requests} for communication with the CDB API. It allows users to list, create and modify
global tags, as well as to inspect their contents. It can be used to download a
global tag for use with the local database backend and to upload a
previously prepared and tested local database configuration to a global tag.

%% file: geometry.tex
In Belle~II, we use the same detailed geometry description for simulation and reconstruction. It is implemented using the Geant4 geometry primitives.
A central service is responsible for setting up the complete geometry: each sub-detector registers a creator that is responsible for defining and configuring its detector-specific volumes as one top-level component of the geometry.

All parameters for the geometry description are provided by payloads in the conditions database.
For the creation of these payloads, a special operation mode is available that reads the geometry parameters from an XML file using libxml2~\cite{libxml2}.
The sub-detector specific descriptions are joined from XML files in the detector packages using XInclude~\cite{XInclude} directives.
The loading from XML includes automatic unit conversion of values that have a ``unit'' attribute and accommodates the definition of new materials and their properties.

Instead of using the conditions database, the geometry can be created directly from XML.
This allows one to edit the XML files to adapt the geometry description as necessary and test the changes locally before creating the payloads and uploading them to the database.

\subsubsection{Testing the Geometry Description}

Developing a functional material and geometry description is quite cumbersome, because, usually, complex construction drawings need to be converted from CAD or paper into code that places the separate volumes with their correct transformation.
To assist the sub-detector developers with this task, we developed a set of tools to supplement the visualization tools provided by Geant4.

First, we run an automated overlap check that uses methods provided by Geant4 to check, for each volume, if it has intersections with any of its siblings or its parent.
This is done by randomly creating points on the surface of the inspected volume and checking if this point is either outside the parent, or inside any of the siblings.
This check is performed on a nightly basis and repeated with more samples points prior to major releases, or if large changes to the geometry have been made.

Second, we provide a module to scan the material budget encountered when passing through the detector.
This module tracks non-interacting, neutral particles through the detector, and records the amount of material encountered along the way.
It can be configured to scan the material in spherical coordinates, in a two-dimensional grid, or as a function of the depth along rays in a certain direction.
The output is a ROOT file containing histograms of the traversed material.
These histograms can be created for each material or each detector component.
In particular, the material distribution by component is a very useful tool to track changes to the material description, allowing us to visualize the differences after each update to the volume-definition code or material-description parameters.

\subsubsection{Magnetic Field Description}

The magnetic field description for Belle~II is loaded from the conditions database.
The payload is created from an XML file using the same procedure as for the geometry description introduced above.
Because the magnetic field does not create any Geant4 volumes, analysis jobs can obtain the field values without the need to instantiate a Geant4 geometry.

The magnetic field creator can handle a list of field definitions for different regions of the detector.
If more than one definition is valid for a given region, either the sum of all field values is taken, or only one definition's value is returned, if it is declared as exclusive.
We have implementations for constant magnetic field, 2D radial symmetric field map and full 3D field maps and some special implementations to recreate the accelerator-magnet conditions close to the beams.
For normal simulation and analysis jobs, we have a segmented 3D field map with a fine grid in the inner-detector region and a total of three coarse outer grids for the two endcaps and the outer-barrel region.

%% file: basf2.bbl
\begin{thebibliography}{10}
\providecommand{\url}[1]{{#1}}
\providecommand{\urlprefix}{URL }
\expandafter\ifx\csname urlstyle\endcsname\relax
  \providecommand{\doi}[1]{DOI \discretionary{}{}{}#1}\else
  \providecommand{\doi}{DOI \discretionary{}{}{}\begingroup
  \urlstyle{rm}\Url}\fi

\bibitem{b2tdr}
T.~{Abe}, et~al., {Belle II Technical Design Report} (2010).
\newblock KEK Report 2010-1

\bibitem{b2tip}
E.~Kou, et~al., {The Belle II Physics book} (2018).
\newblock KEK Preprint 2018-27

\bibitem{cvmfs}
{CernVM File System}.
\newblock \urlprefix\url{https://cernvm.cern.ch/portal/filesystem}.
\newblock Accessed 26.11.2018

\bibitem{root}
R.~Brun, F.~Rademakers, Nucl. Instrum. Meth. \textbf{A389}(1), 81 (1997).
\newblock \doi{10.1016/S0168-9002(97)00048-X}

\bibitem{geant4}
S.~Agostinelli, et~al., Nucl. Instrum. Meth. \textbf{A506}, 250 (2003).
\newblock \doi{10.1016/S0168-9002(03)01368-8}

\bibitem{evtgen}
D.~Lange, Nucl. Instrum. Meth. \textbf{A462}, 152 (2001).
\newblock \doi{10.1016/S0168-9002(01)00089-4}

\bibitem{spack}
Spack.
\newblock \urlprefix\url{https://spack.io/}.
\newblock Accessed 26.11.2018

\bibitem{bitbucket}
Bitbucket.
\newblock \urlprefix\url{https://bitbucket.org}.
\newblock Accessed 26.11.2018

\bibitem{jira}
Jira.
\newblock \urlprefix\url{https://www.atlassian.com/software/jira}.
\newblock Accessed 26.11.2018

\bibitem{scons}
{SCons -- A software construction tool}.
\newblock \urlprefix\url{http://www.scons.org/}.
\newblock Accessed 26.11.2018

\bibitem{astyle}
{AStyle}.
\newblock \urlprefix\url{http://astyle.sourceforge.net/}.
\newblock Accessed 26.11.2018

\bibitem{pep8}
{pep8 - Python style guide checker}.
\newblock \urlprefix\url{http://pep8.readthedocs.io}.
\newblock Accessed 26.11.2018

\bibitem{autopep8}
autopep8.
\newblock \urlprefix\url{https://github.com/hhatto/autopep8}.
\newblock Accessed 26.11.2018

\bibitem{googletest}
{Google {C++} Testing Framework}.
\newblock \urlprefix\url{https://code.google.com/p/googletest/}.
\newblock Accessed 26.11.2018

\bibitem{bamboo}
Bamboo.
\newblock \urlprefix\url{https://www.atlassian.com/software/bamboo}.
\newblock Accessed 26.11.2018

\bibitem{buildbot}
Buildbot.
\newblock \urlprefix\url{https://buildbot.net/}.
\newblock Accessed 26.11.2018

\bibitem{sphinx}
{SPHINX Python Documentation Generator}.
\newblock \urlprefix\url{http://www.sphinx-doc.org}.
\newblock Accessed 26.11.2018

\bibitem{genfit}
C.~Höppner, S.~Neubert, B.~Ketzer, S.~Paul, Nucl. Instrum. Meth.
  \textbf{A620}, 518 (2010).
\newblock \doi{10.1016/j.nima.2010.03.136}

\bibitem{genfit2}
{Rauch, Johannes and Schl\"uter, Tobias}, J. Phys. Conf. Ser. \textbf{608}(1),
  012042 (2015).
\newblock \doi{10.1088/1742-6596/608/1/012042}

\bibitem{boost.python}
{Boost.Python}.
\newblock
  \urlprefix\url{https://www.boost.org/doc/libs/1_64_0/libs/python/doc/html/index.html}.
\newblock Accessed 26.11.2018

\bibitem{pyroot}
{PyROOT}.
\newblock \urlprefix\url{https://root.cern.ch/pyroot}.
\newblock Accessed 26.11.2018

\bibitem{ipython}
F.~Perez, B.E. Granger, Comput. Sci. Eng. \textbf{9}(3), 21 (2007).
\newblock \doi{10.1109/MCSE.2007.53}

\bibitem{jupyter}
Jupyter.
\newblock \urlprefix\url{http://jupyter.org/}.
\newblock Accessed 26.11.2018

\bibitem{matplotlib}
J.D. Hunter, Comput. Sci. Eng. \textbf{9}(3), 90 (2007).
\newblock \doi{10.1109/MCSE.2007.55}

\bibitem{pandas}
{Python Data Analysis Library}.
\newblock \urlprefix\url{https://pandas.pydata.org/}.
\newblock Accessed 26.11.2018

\bibitem{hep-ipython-tools}
{HEP IPython Tools}.
\newblock \urlprefix\url{http://hep-ipython-tools.github.io/}.
\newblock Accessed 26.11.2018

\bibitem{zeromq}
{{\O}MQ}.
\newblock \urlprefix\url{http://zeromq.org/}.
\newblock Accessed 26.11.2018

\bibitem{DBLP:journals/corr/Vigna14}
S.~Vigna, CoRR \textbf{abs/1402.6246} (2014).
\newblock \urlprefix\url{http://arxiv.org/abs/1402.6246}

\bibitem{Marsaglia:2003:JSSOBK:v08i14}
G.~Marsaglia, Journal of Statistical Software \textbf{8}(14), 1 (2003).
\newblock \urlprefix\url{http://www.jstatsoft.org/v08/i14}

\bibitem{sha3}
Q.H. Dang, Federal Inf. Process. Stds.  (2015).
\newblock \doi{10.6028/NIST.FIPS.202}

\bibitem{conditions-db}
L.~Wood, T.~Elsethagen, M.~Schram, E.~Stephan, Journal of Physics: Conference
  Series \textbf{898}(4), 042060 (2017).
\newblock \urlprefix\url{http://stacks.iop.org/1742-6596/898/i=4/a=042060}

\bibitem{python-requests}
{Requests: HTTP for Humans}.
\newblock \urlprefix\url{http://docs.python-requests.org/}.
\newblock Accessed 26.11.2018

\bibitem{libxml2}
{The XML C parser and toolkit of Gnome}.
\newblock \urlprefix\url{http://xmlsoft.org/}.
\newblock Accessed 26.11.2018

\bibitem{XInclude}
{XML Inclusions}.
\newblock \urlprefix\url{https://www.w3.org/TR/xinclude/}.
\newblock Accessed 26.11.2018

\end{thebibliography}
